\documentclass[preprint,authoryear,review,12pt]{elsarticle}

\usepackage{amssymb}
\usepackage{amsmath}
\usepackage{epstopdf}
\usepackage{lineno}

\journal{Ocean Modelling}

\begin{document}

\begin{frontmatter}



\title{\textbf{\large{Calibration of the K-Profile Parameterization of ocean boundary layer mixing.  Part I: Development 
of an inquiry dependent test statistic 
}}}


\author[i1,i2]{Sarah E. Zedler}
\ead{szedler@ig.utexas.edu}
\fntext[]{Corresponding author phone number: 512-471-0389}
\author[i2]{Charles S. Jackson}
\author[i1]{Fengchao Yao}
\author[i3]{Patrick Heimbach}
\author[i4]{Armin K\"ohl}
\author[i5]{Rob B. Scott}
\author[i1]{Ibrahim Hoteit}

\address[i1]{King Abdullah University of Science and Technology, Thuwal, Saudi Arabia}
\address[i2]{Institute for Geophysics, University of Texas at Austin, Austin, TX, USA}
\address[i3]{Massachusetts Institute of Technology, Boston, MS, USA}
\address[i4]{Universit\"at Hamburg, Hamburg, Germany}
\address[i5]{Laboratoire de Physique des Oceans, CNRS, University of Brest, France}
\begin{abstract}
In model comparisons with observational data, not all data contain information that is useful for answering a specific science question.  If non-relevant or highly uncertain data are included in a comparison metric, they can reduce the significance of other observations that matter for the scientific process of interest.  Sources of noise and correlations among summed quantities within a comparison metric  affect the significance of a signal that is useful for testing model skill. In the setting of the tropical Pacific, we introduce an ``inquiry dependent'' (ID) metric of model-data comparison that determines the relative importance of the TOGA-TAO buoy observations of the ocean temperature, salinity, and horizontal currents for influencing upper-ocean vertical turbulent mixing as represented by the K-Profile Parameterization (KPP) embedded in the MIT general circulation model (MITgcm) for the 2004-2007 time period.  The ID metric addresses a challenge that the wind forcing is likely a more significant source of uncertainty for the ocean state than the turbulence itself, and that the observations are correlated in time, space, and across ocean state variables.  In this approach the MITgcm is used to infer variability and relationships in and among the data, and to determine the response structures that are most relevant for constraining uncertain parameters.   We demonstrate that the ID metric is able to distinguish the effects due to parameter perturbations from those due to uncertain winds and that it is important to include multiple kinds of data in the comparison, suggesting that the ID metric is appropriate for use in a calibration of the KPP model parameters using mooring observations of the ocean state.  
\end{abstract}

\begin{keyword}

%% MSC codes here, in the form: \MS

\end{keyword}

\end{frontmatter}
%\begin{linenumbers}
\section{Introduction}
In uncertainty quantification (UQ) the goal is to calibrate model parameters using observations.  This requires having a way to compare models and data.  However, the UQ community has widely ignored details concerning how complex models are compared to data.  When it comes to uncertainty quantification of systems of sufficient complexity, there is a significant role for scientific understanding of the processes, data, and sources of uncertainty that can affect the success of our ability to make use of the data to constrain uncertain parameters.  For large scale models, there remains a sizable ``irreducible'' error \citep{mcw07}. It is not clear how well we can calibrate such models using observational data if the end  result is getting matches to data for the wrong reasons.  One goal of this paper is to introduce a method of making model-data comparisons that takes these considerations into account and to provide some level of understanding of how the data would be used to test model physics.

We are interested in how observations of the ocean state in the tropical Pacific collected from the TOGA-TAO moored buoy array \citep{mcp98} can be used to calibrate parameters within the K-Profile parameterization of turbulent mixing \citep{lar94} as embedded in a regional version of the the MIT general circulation model \citep{adc95,mar97a,mar97b}.  The coupling between the atmosphere and ocean and the behavior of the ENSO is very sensitive to ocean mixing processes.   While mixing events take place over a matter of hours, the effects of mixing will integrate much longer time scales of months to 100 years or more.  Moreover while we observe the effects of mixing events at discrete points, there are significant uncertainties in inferences of surface wind stresses.   So while we are interested in correctly representing the short and long term effects of boundary layer mixing, we are also challenged by the shortness of the observational record, uncertainties in forcing, and the chaotic nature of variability. 

Most uncertainty quantification investigations use a simple metric for model-data comparison that is just a sum over squared model-data differences that have been normalized on a point-by-point basis by the variance in the data.  This simple metric is often used within state estimation or inverse modeling, which is an optimization problem with the goal of finding the parameter values, initial values, or boundary conditions that allow a model to best approximate observations.  The simple metric is less appropriate when the goal is to solve for the uncertainty in the optimal solution, which depends on strength of the observational evidence to accept or reject alternate solutions. This uncertainty will be improperly represented by the simple test statistic if there are correlations in the data or if the data are of varying quality.
 
In development of our ideas of how to compare models to data, we will discuss three main questions within the framework of our turbulent mixing application:
\begin{itemize}
\item Is there a way to get around the uncertainty contributed by the wind?
\item How important is it that we synthesize all available data?
\item What is the potential to make use of moored buoy observations for calibrating uncertain parameters within the KPP?
\end{itemize}

\section{``Inquiry Dependent'' test statistic methodology}
In model data comparisons, one needs to make use of data that are relevant to the question one is asking.  Data will be affected by many processes including those that are unrelated to one's interests and by uncertain initial conditions and forcing.  In order to focus on data relevant to parametric uncertainties, we have developed an Inquiry Dependent (ID) filter that makes use of an ensemble of parameter sensitivity experiments as well as a way of generating the effects of uncertain initial and forcing conditions. 

\subsection{Filtering for signal of interest}
Assuming modeling errors follow a multivariate normal distribution, the likelihood of using model $g(\mathbf{m})_v$ with parameters $\mathbf{m}$ to simulate $k$ observations $\mathbf{d}_v$ of ocean state variable $v$ is given by
\begin{equation}
\theta(\mathbf{d}_v|g(\mathbf{m})_v)=\frac{1}{\sqrt{(2\pi)^{k} | \mathbf{C}_v |}} exp\Big(-\frac{1}{2}(g(\mathbf{m})_v-\mathbf{d}_v)^T\mathbf{C}^{-1}_v(g(\mathbf{m})_v-\mathbf{d}_v)\Big),
\label{Eq:matrix_cost_k}
\end{equation}
which only works mathematically if the rank of covariance matrix $\mathbf{C}_v$ is equal to the number of observations $k$. If not, then the inverse of the covariance matrix is singular and its determinant is $0$. Such is the case with this problem since the dynamics of the atmosphere-ocean system creates covariance structures which dramatically reduce the system's effective degrees of freedom. A solution to this problem is to apply singular value decomposition to $\mathbf{C}_v$ in order to identify a limited set of $k_e<k$ eigenvectors associated with its largest eigenvalues. These eigenvectors are commonly referred to as empirical orthogonal functions or `EOFs' within the climate literature \citep[e.g.,][]{muq04}. The argument of the exponent in equation (\ref{Eq:matrix_cost_k}), when rotated into this new orthogonal basis and truncated to include the first $k_e$ vectors associated with the largest eigenvalues $\lambda$, is equivalent to a $\chi^2_{(k_e)}$ test statistic with $k_e$ degrees of freedom. We refer to this metric as a `cost' function $E(\mathbf{m})_v$ for component $v$ and is given by
\begin{equation}
E(\mathbf{m})_v= \sum_{i=1}^{k_e}\frac{[\mathbf{eof}_{i,v}^T(g(\mathbf{m})_v-\mathbf{d}_v)]^2}{2\lambda_{i,v}} = \frac{1}{2}\chi^2_{(k_e)}.
\label{chi-squared}
\end{equation}

The idea of the ID test statistic is to select EOFs related to a covariance matrix constructed from changes in model parameters $\mathbf{m}$ on each field $v$. These ID specific EOFs filter the observational data for the structures related to parameters, which are hopefully, but are not necessarily, distinct from sources of uncertainty in initial conditions and the wind forcing. The null hypothesis still needs to be represented by the effect of winds and internal variability on observables $\mathbf{d}_v$. We therefore do not use the eigenvalues associated with the SVD decomposition of $\mathbf{C}_v$ generated from parametric sensitivity experiments. Instead we estimate variances $\lambda_{i,v}$ in equation \ref{chi-squared} from $N_{exp}$ experiments that test the effect of uncertain winds and initial conditions on each ID EOF amplitude, 
\begin{equation}
\lambda_{i,v} = var(\mathbf{eof}_{i,v}^T[g(\mathbf{m})_{v,j}-\overline{g(\mathbf{m})_{v,j}}]), j=1,2, ... N_{exp}.
\label{var}
\end{equation}
The parametric sensitivities used to generate the ID specific EOFs and simulations representing uncertainties in initial conditions and wind forcing is reviewed in Section \ref{Experiments}.

\subsection{Cost component weighting}
The ID test statistic represented by equation (\ref{chi-squared}) can be obtained for different time segments and fields. Summing up components assumes the components are independent whereas averaging assumes the components are dependent. Neither is likely entirely correct. \citet{jac16} provides a strategy within a multivariate normal framework for assigning weights to individual components and their sum. The process makes use of a set of idealized model experiments where model output is used as observational data. From this ensemble one may estimate both the effective degrees of freedom $\{k_e\}_v$ as well as component weights $<S_v>$. Further down we justify setting a lower cutoff value for $k_{cut}=7<k_e$, which is the same for all cost function components, based on ID metric goals outlined in the results section. The expression for the total cost function including a weighted sum over its four components, one each for temperature, salinity, zonal `u' currents, and meridional `v' currents, is given by
\begin{equation}
E(\mathbf{m})_{tot}= S_{tot}\sum_{v=1}^{4}<S_v>\frac{1}{N_t}\sum_{t=1}^{N_t}\sum_{i=1}^{k_{cut}}\frac{[\mathbf{eof}_{i,v}^T(g(\mathbf{m},t)_v-\mathbf{d}(t)_v)]^2}{2\lambda_{i,v}}.
\label{Etotal}
\end{equation}
Table (\ref{tab:S}) provides the values we used for $<S_v>$ in equation(\ref{Etotal})
\begin{table}[tp]
\begin{tabular}[t]{|c|c|c|}
\hline
Component & $<S_q>$ & $k_e$\\ \hline 
$S_{uvel}$ & 8.72 & 8\\ \hline
$S_{vvel}$ & 3.55  & 16\\ \hline 
$S_{temp}$ & 9.06 & 8\\ \hline 
$S_{salt}$ & 6.24 & 12\\ \hline 
$S_{tot}$ & 0.29 & 17\\ \hline 
\end{tabular}
\caption{List of scaling factors used to weight cost function components. The  Here \textit{uvel} is east velocity, \textit{vvel} is north velocity, \textit{temp} is temperature and \textit{salt} is salinity. $k_e$ is the estimated degrees of freedom.   
}
\label{tab:S}
\end{table}

\section{Model and Data}
We use a version of the Massachusetts Institute of Technology general circulation model (MITgcm) implemented with the K-Profile Parameterization turbulent mixing scheme \citep{adc95,mar97b,mar97a,lar94} in a regional configuration
based on that of \citet{hot08} and \citet{hot10} to simulate the ocean flow.  The KPP has been described in detail elsewhere \citep{lar94,lar99}. We present a basic description of some of the key features and how they relate to the constant parameters we have perturbed here.  
\subsection{MITgcm Model}
\begin{figure}
\begin{center}
\includegraphics[width=\textwidth,height=3.0in]{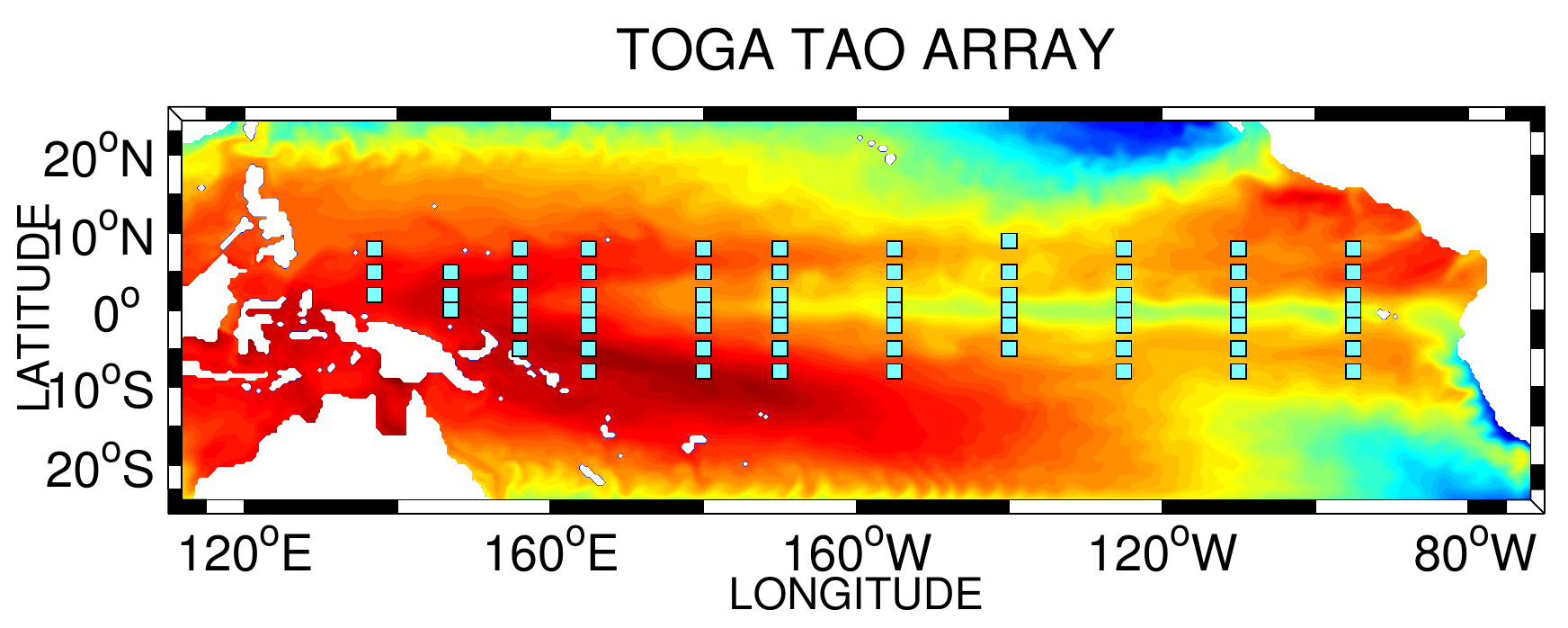}
\caption{
        Model domain showing 90-day averaged sea surface temperature for March-May 2007 and the TOGA/TAO mooring array.  
        }
\label{setup}
\end{center}
\end{figure}

The MITgcm is based on the primitive Navier Stokes equations.  The model is implemented with an implicit non-linear free surface in spherical coordinates.  The bi-harmonic background viscosity and diffusivity have values of $4.0\times10^{11}~m^4/s$ and $2.0\times10^{11}~m^4/s$, respectively.  The horizontal resolution is $1/3^\circ$ for a domain covering the
tropical Pacific region with an extent from -26$^o$N to 30$^o$N, and 104$^o$E to 70$^o$W (Fig. \ref{setup}). The vertical resolution
%%%%%%%%%%%%%%%%%%%%%%%%%%%%%%%%%%%%%%%%%%%%%%%%%%%%%%%%%%%%%%%%%%%%%%%%%%%%%%%%%%%%%%%%%%%%%%%%%%%%%%%%%%%%%%%%%%%%%%%%%%
%%%%%%%%%%%%%%%%%%%%%%%%%%%%%%%%%%%%%%%%%%%%%%%%%%%%%%%%%%%%%%%%%%%%%%%%%%%%%%%%%%%%%%%%%%%%%%%%%%%%%%%%%%%%%%%%%%%%%%%%%%
ranges between 5 m near the surface to 510 m at the bottom, with layer thicknesses increasing logarithmically beneath 71m depth (51 layers total). The ETOPO2v2 topography available at 2 minute resolution was interpolated onto our coarser grid.  The model timestep is 900 seconds.  The initial conditions for temperature, salinity, and velocity components come from the OCean Comprehensible Atlas \citep[OCCA;][]{for10} reanalysis product that is based on the MITgcm and most available ocean data sets for our time period of interest, 2003-2007 (starting on November 1).  The model has a sponge layer (with a thickness of 9 grid cells and inner and outer boundary relaxation timescales of 20 and 1 days respectively) and open lateral boundaries, which are extracted from the OCCA data assimilation product and interpolated at the model resolution of $1/3^o$ and have a time step of one day.  Therefore, at the initial timestep, the boundary conditions are in equilibrium with the interior fields. The transport in the Indonesian throughflow is adjusted at each timestep so that there is no net flow into or out of the model domain.  

Unless otherwise specified, the model is forced with surface heat and momentum fluxes calculated in the MITgcm from the nominally $1.8^\circ$ and six-hourly NCEP/NCAR Reanalysis atmospheric state data for the period 2003-2007 \citep{kal96}.  The wind speed is then converted to wind stress using the drag coefficient relationship of \citet{lar04} and the bulk formula for wind stress \citep{gil82}.  The heat fluxes and wind stress are restorative, in the sense that they depend on the surface current speed and/or sea surface temperature as diagnosed at each time step of the MITgcm. 

In three experiments, the wind velocity is set from either the ECMWF \citep{gib97}, NOAA/CIRES Twentieth Century reanalyses \citep{com11}, or NASA Cross-Calibrated Multi-Platform Ocean Surface Wind Velocity product \citep{atl96}. For assessing uncertainty caused by wind forcing, an additional 20 experiments are forced with a linear combination of NCEP/NCAR, ECMWF, and NASA products.  The reanalysis products represent different estimates of the real wind field.
\subsection{KPP}
A detailed description of the KPP model is provided in \citet{lar94}.  Here, we provide a brief summary of some of the key features of the model.  A summary of the basic function of the nine parameters in the KPP model is presented in Table \ref{function}.
%%%%%%%%%%%%%%%%%%%%%%%%%%%%%%%%%%%%%%%%%%%%%%%%%%%%%%%%%%%%%%%%%%%%%%%%%%%%%%%%%%%%%%%%%%%%%%%%%%%%%%%%%%%%%%%%%%%%%%%%%%
\begin{table}[tp]
\scalebox{0.5}{
\begin{tabular}[t]{|c|c|c|}
\hline
\multicolumn{3}{|c|}{KPP Parameters Varied and Basic Function} \\ \hline
Parameter Name & Symbol & Basic Function \\ \hline
critical bulk Richardson number & $Ri_c$ & affects depth and magnitude of $K_x$ in BL under convective and shear driven forcing \\ \hline
critical gradient Richardson number & $Ri_c$ & affects depth and magnitude of $K_x$ in interior under convective and shear driven forcing \\ \hline
structure function, stable forcing & $\phi_{stbl}$ & affects magnitude of $K_x$ in BL under stable forcing \\ \hline
structure function, momentum, unstable forcing & $\phi_{m,unst}$ & affects magnitude of $K_u$ in BL under convective forcing \\ \hline
structure function, temperature or salinity, unstable forcing & $\phi_{m,unst}$ & affects magnitude of $K_s$ in BL under convective forcing \\ \hline
non-local convective forcing parameter & $C^*$ & affects profile of turbulent fluxes in BL under convective forcing conditions \\ \hline
interior viscosity due to shear &  $\nu^s$ & scales magnitude of $K_u$ in interior \\ \hline
brunt-Vaisala frequency for convection & $N_0^2$ & modulates magnitude of $K_s$ in interior \\ \hline
interior diffusivity for convection & $\nu^c$ & scales magnitude of $K_s$ in interior \\ \hline
\end{tabular}
}
\caption{List of nine parameters varied in this experiment along with the symbols used to refer to them and a brief description of their
basic function. In this table, BL stands for the KPP diagnosed Boundary Layer.}
\label{function}
\end{table}
%%%%%%%%%%%%%%%%%%%%%%%%%%%%%%%%%%%%%%%%%%%%%%%%%%%%%%%%%%%%%%%%%%%%%%%%%%%%%%%%%%%%%%%%%%%%%%%%%%%%%%%%%%%%%%%%%%%%%%%%%%

Overall, the MITgcm model calculates turbulent flux profiles following
\begin{equation}
\overline{wx}(d)= -K_x(\frac{\partial X}{\partial z} - \gamma_x)
\label{wx}
\end{equation}
where $x=s$ for properties of temperature or salinity, and $x=u$ for properties of horizontal current components, $K_x$ is the eddy diffusivity or viscosity, and $\gamma_x$ (which is proportional to the constant parameter $C^*$) is a term that represents the effect of surface forced convective mixing on the turbulent fluxes in the ocean turbulent boundary layer.  The KPP calculates the vertical profile of $K_x$ at each coordinate in the domain and at each timestep.

For the parameterization of $K_x$, the KPP recognizes two types of turbulent mixing: convection and
Kelvin-Helmholz shear instabilities.  As such, two important non-dimensional scales in the KPP are the bulk and gradient Richardson numbers.  In general terms, both of these scales are a measure of how turbulent the water column is, and are defined as the ratio of density stratification to squared vertical shear in the horizontal currents.  Negative Richardson numbers indicate that there is net cooling at the surface, inducing a density inversion that results in vigorous turbulent mixing.  Small positive values indicate that although there is net warming at the surface (stable surface forcing), there is sufficient shear in the horizontal currents to induce Kelvin-Helmholz type instabilities.   The bulk and gradient Richardson numbers differ in the depth regions over which the stratification and shear derivatives are calculated.  The bulk Richardson number is based on first order differences between a near-surface value and the depth of interest.  The gradient Richardson number is based on derivatives calculated at the vertical grid scale between adjacent depth levels.  

In the KPP, the water column is divided into two depth regions with separate prescriptions for $K_x$.  The boundary layer 
depth $h$ that separates them is defined as the shallowest bulk Richardson number, which typically increases with depth, that is equal to a constant critical value, $Ri_c$.  In the boundary layer, $K_x$ is proportional to the product of $h$ and 
a vertical velocity scale that is proportional to strength of the wind forcing (through the friction velocity) and has separate parameterizations for stable and unstable surface heat flux forcing conditions (i.e. whether the net heat flux is going into or out of the ocean).  Here the relevant parameters are $\phi_{stbl}$, $\phi_{m,unst}$, $\phi_{s,unst}$ and $N^2_0$. In the interior, the eddy diffusivity is a smoothly decreasing function of increasing gradient Richardson number over a depth region constrained by a critical value $Ri_g$ and is proportional to a maximum value ($\upsilon^c$ for tracers and $\upsilon_s$ for momentum). After all of these quantities have been diagnosed, the boundary layer diffusivities are scaled through multiplication by a bell-shaped polynomial with depth that takes a maximum at mid-depth, requires the final $K_x$ profile to be continuous and smooth across $h$, and that tapers smoothly to zero at the surface and the depth of the interior. 
%%%%%%%%%%%%%%%%%%%%%%%%%%%%%%%%%%%%%%%%%%%%%%%%%%%%%%%%%%%%%%%%%%%%%%%%%%%%%%%%%%%%%%%%%%%%%%%%%%%%%%%%%%%%%%%%%%%%%%%%%%
\begin{table}[tp]
\scalebox{0.6}{
\begin{tabular}[t]{|c|l|c|l|c|c|}
\hline
Expt. No. & Parameter Name or Wind Stress Type & Symbol & Perturbed Value & Default Value & KPP Eq. No. \\ \hline
1 & ECMWF Wind & $N/A$ & $N/A$ & NCEP/NCAR &  $N/A$ \\ \hline
2 & NOAA CIRES Wind & $N/A$ & $N/A$ & NCEP/NCAR& $N/A$ \\ \hline
3 & NASA Wind & $N/A$ & $N/A$ & NCEP/NCAR& $N/A$ \\ \hline
4 & critical bulk Richardson number & $Ri_c$ & $Ri_c$=0.15 & $Ri_c$=0.30 & $Ri_b$, 21 \\ \hline
5 & critical bulk Richardson number & $Ri_c$ & $Ri_c$=0.45 & $Ri_c$=0.30 & $Ri_b$, 21 \\ \hline
6 & critical bulk Richardson number & $Ri_c$ & $Ri_c$=0.60 & $Ri_c$=0.30 & $Ri_b$, 21 \\ \hline
7 & critical gradient Richardson number & $Ri_g$ & $Ri_g$=0.1 & $Ri_g$=0.7 & $Ri_g$, 27\\\hline
8 & critical gradient Richardson number & $Ri_g$ & $Ri_g$=1.0 & $Ri_g$=0.7 & $Ri_g$, 27 \\\hline
9 & structure function, stable forcing & $\phi_{stbl}$ & $b_{m,s}$=8.0 & $b_{m,s}$=5.0 & constant in B1a \\\hline
10 & structure function, stable forcing & $\phi_{stbl}$ & $b_{m,s}$=2.0 & $b_{m,s}$=5.0 & constant in B1a\\\hline
11 & structure function, unstable forcing, momentum & $\phi_{m,unst}$ & $b_{m,unst}$=331.06 & $b_{m,unst}$=16 & constant in B1b\\\hline
12 & structure function, unstable forcing, momentum & $\phi_{m,unst}$ & $b_{m,unst}$=3.60 & $b_{m,unst}$=16 & constant in B1b\\\hline
13 & structure function, unstable forcing, tracer & $\phi_{s,unst}$ & $b_{s,unst}$=67.02 & $b_{s,unst}i$=16 & constant in B1d\\\hline
14 & structure function, unstable forcing, tracer & $\phi_{s,unst}$ & $b_{s,unst}$=7.77 & $b_{s,unst}=16$ & constant in B1d\\\hline
15 & nonlocal transport & $\gamma_s$ & $C^*$=5.0 & $C^*$=10.0 & $C^*$, 19\\\hline
16 & nonlocal transport & $\gamma_s$ & $C^*$=15.0 & $C^*$=10.0& $C^*$, 19\\\hline
17 & interior viscosity due to shear & $\upsilon^s$ & $\upsilon^0$=25$\times 10^{-4}$ m$^2$s$^{-1}$ & $\upsilon^0=50\times 10^{-4}$ m$^2$s$^{-1}$&$\upsilon^0$, 28\\\hline
18 & interior viscosity due to shear & $\upsilon^s$ & $\upsilon^0$=75$\times 10^{-4}$ m$^2$s$^{-1}$ & $\upsilon^0=50\times 10^{-4}$ m$^2$s$^{-1}$&$\upsilon^0$, 28 \\\hline
19 & brunt-Vaisala frequency for convection & $N^2_0$ & $N^2_0$=-0.1$\times 10^{-4}$ s$^{-1}$ & $N^2_0$=-0.2 $\times 10^{-4}$ s$^{-1}$ & not shown, related to 28 \\\hline
20 & brunt-Vaisala frequency for convection & $N^2_0$ & $N^2_0$=-0.3$\times 10^{-4}$ s$^{-1}$ & $N^2_0$=-0.2 $\times 10^{-4}$ s$^{-1}$ & not shown, related to 28\\\hline
21 & interior diffusivity due to convection & $\upsilon^c$ & $\upsilon^{0,c}$=0.05 m$^2$s$^{-1}$ &$\upsilon_{0,c}$=0.10 m$^2$s$^{-1}$ & like $\upsilon^0$ for conv., 28\\\hline
22 & interior diffusivity due to convection & $\upsilon^c$ & $\upsilon^{0,c}$=0.15 m$^2$s$^{-1}$  &$\upsilon_{0,c}$=0.10 m$^2$s$^{-1}$ & like $\upsilon^0$ for conv., 28\\\hline
\end{tabular}
}
\caption{List of experiments for single parameter perturbation and pure wind cases, as well as the default values for the KPP as implemented in the MITgcm. Units of parameter settings are noted; the lack of notation indicates the parameter is unitless.  The ``Symbol" column provides the symbol we use to refer to the ``Parameter Name" in the text.  We have tried to stay as close to the KPP notation as possible \citep{lar94}.  ``Perturbed Value" is the value of the perturbed parameter set in the indicated sensitivity experiment. ``KPP Eq. No." lists the equivalent symbol used in \citet{lar94} followed by the equation number that the variable is introduced.  Note that \citet{lar94} does not show the exact equations for variables perturbed in Experiments 19-22. }
\label{tab:linexpt}
\end{table}
%%%%%%%%%%%%%%%%%%%%%%%%%%%%%%%%%%%%%%%%%%%%%%%%%%%%%%%%%%%%%%%%%%%%%%%%%%%%%%%%%%%%%%%%%%%%%%%%%%%%%%%%%%%%%%%%%%%%%%%%%%
\begin{figure}[!h]
\begin{center}
\includegraphics[width=\textwidth,height=7.0in]{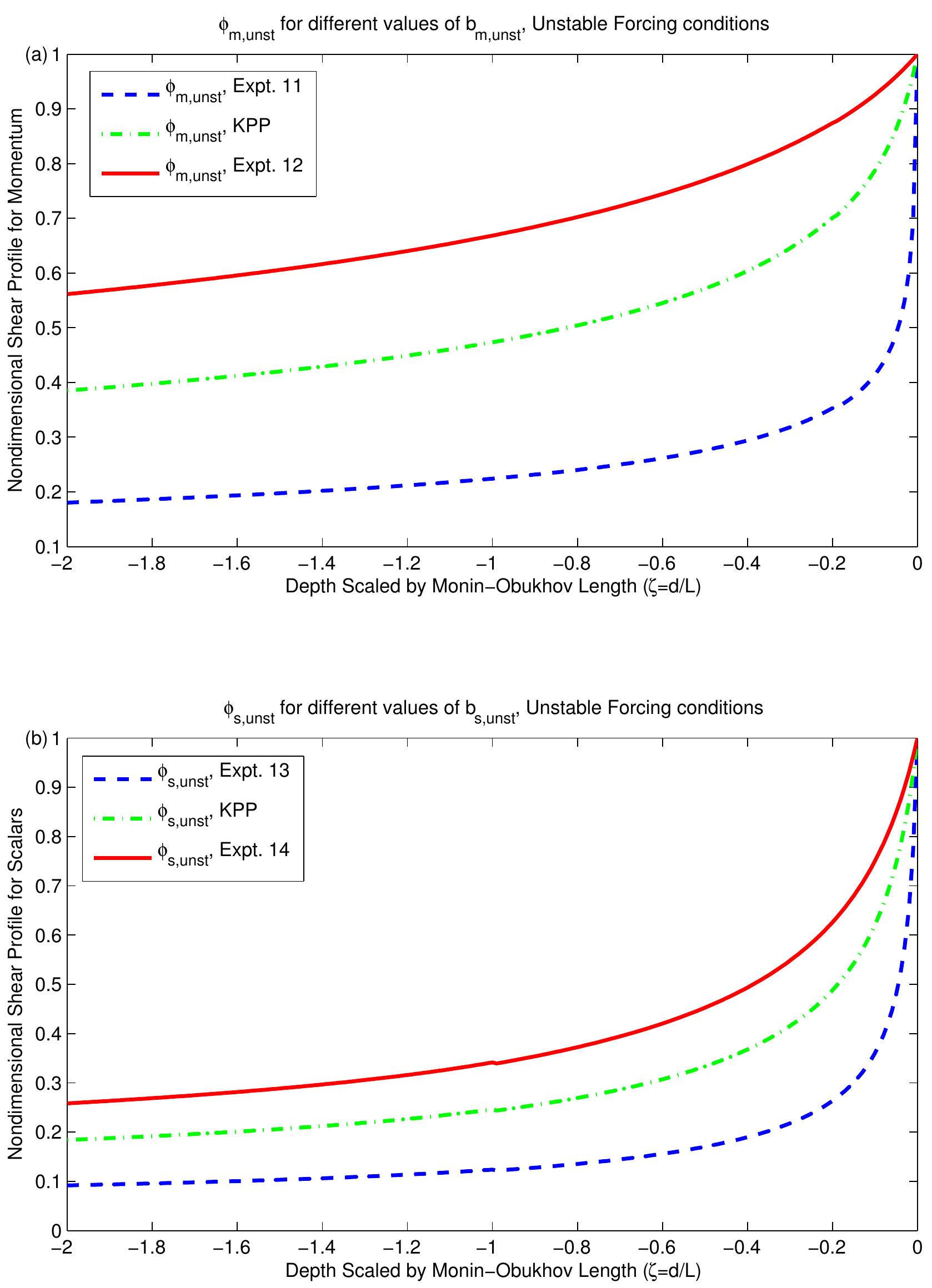}
\caption{
(a) Non-dimensional shear profiles for momentum as a function of scaled depth ($\phi_{m,unst}$), as specified for the default case (KPP), experiment 11, and experiment 12.  (b) Non-dimensional shear profiles for scalars (tracers) as a function of scaled depth ($\phi_{s,unst}$), as specified for the default case (KPP), as well as perturbed experiments 13 and 14. }
\label{plot_phi}
\end{center}
\end{figure}
%%%%%%%%%%%%%%%%%%%%%%%%%%%%%%%%%%%%%%%%%%%%%%%%%%%%%%%%%%%%%%%%%%%%%%%%%%%%%%%%%%%%%%%%%%%%%%%%%%%%%%%%%%%%%%%%%%%%%%%%%%
\subsection{Data}
The observational data for our experiment come from the TOGA-TAO mooring array (locations shown in Fig. \ref{setup}) 
for the November 2003--November 2007 time period.  The array consists of 77 moorings centered on the Equator that span the width of the tropical Pacific in the east-west direction in the latitude range from 8$^o$S to 8$^o$N \citep{mcp98}.    
In tests of the KPP parameters, all available daily averaged temperature, salinity, and horizontal velocity data at or above 300m were used.  Velocity measurements were made using a combination of current meters at various latitudes and four uplooking Acoustic Doppler Current Profilers (ADCPs) along the equator.   The depth 300m was chosen because deeper 
data would likely not be directly affected by upper ocean turbulent mixing, as parameterized in the KPP.  There were 699 temperature sensors, 145 salinity platforms, and 218 velocity positions.

Before comparing the model to observations, we apply a 90-day average, which is appropriate for synthesis of a coarse resolution model.  We choose this 90-day timescale rather than a diurnal timescale because we are interested in quantifying the uncertainty in KPP parameters for a coarse scale model run for a long time.  The default KPP parameters have been validated for short length- and time-scales \citep{lar94,lar99}. There is no guarantee that a parameterization tuned for short timescales will be appropriate for long timescales. Furthermore, we cannot expect our regional coarse scale model to reproduce individual mixing events on short timescales accurately.  By averaging over 90-days, we attempt to capture the time-integrated or cumulative effect of the diurnal cycle.  The 90-day average also removes high frequency fluctuations in the ocean state caused by eddies and Tropical Instability waves that we do not expect the model to be able to accurately reproduce.  Tropical instability waves have periods of 3-4 weeks \citep{phi86}.

In this study, the data are used to determine how close observed and modeled fields are to one another for a given parameter setting.  This implies that we should wait until the perturbed parameter has a chance to fully affect the simulated ocean state on seasonal timescales before calculating the test statistic.  We choose a model spin-up time of 1.5 years, which is probably not long enough but balances timescales of ocean adjustment and computational demands of the model.  One and a half years is longer than the timescales for Rossby and Kelvin waves (i.e. thermocline anomalies and downwelling), which travel from their generation sites to the Equator and across the Pacific in about 6-9 months \citep{boc04}.  It is also consistent with the timescale of a few years for vertical turbulent mixing and heat exchange to occur in the upper ocean (entrainment) once the thermocline anomalies have been advected around the basin (estimated by \citet{boc05} to be few years).  Completing one 4-year integration of the model requires approximately a day of wall clock time and consumes \~25k cpu hours.

\section{Experiments}
\label{Experiments}
\subsection{Individual Wind and Parameter Experiments}
We consider 19 individual parameter perturbation experiments.  These are associated with mixing in the boundary layer,
 interior mixing, or the non-local convective mixing.  In Table \ref{tab:linexpt} for each experiment, we list the corresponding symbols and equation numbers in \citet{lar94} where individual parameters are introduced.  In addition, we test for the effects of three different wind speed products using the default KPP parameter settings (Exp. 1-3).  Among the list of parameter sensitivity experiments are ones for the critical bulk and gradient Richardson numbers $Ri_c$ (Exp. 4-6) and $Ri_g$ (Exp. 7-8).  Under stable forcing conditions, the vertical velocity scale profile is modified by adjusting a polynomial coefficient, with excursions of a maximum of 50$\%$ and 100$\%$ (Fig. \ref{plot_phi}).  Similar perturbations are made to the non-dimensional shear and stratification profiles for momentum (Exp. 11-12; $\phi_{m,unst}$), and temperature and salinity (Exp 13-14; $\phi_{s,unst}$).  The non-local transport term $C^*$, which has to do with convective instabilities, 
is perturbed by 50$\%$ \citep[Exp. 15 and 16;][]{mai82}.  In the interior, the maximum $K_u$  (Exp. 17-18; $\upsilon^s$) and $K_s$ (Exp. 20-21; $\upsilon^c$) are modified by 50$\%$ higher and lower than the default.  Another threshold that is associated with $\upsilon^s$ for convection, $N_0^2$, is perturbed by 100$\%$ (Exp. 19 and 20).  We choose to probe the sensitivity of the model by making the perturbations large.  If there is no sensitivity in the model to parameter perturbations, then it is an indication that the parameter is not as important for setting mixing in the upper ocean, at least in the fashion it is being incorporated into the test statistic.  In that case, attempting to use a model and data to constrain ocean mixing parameters would be a futile exercise. 
\subsection{Blended Wind Experiments} 
Experiments W1 through W20 are generated with the default KPP parameters, but in each case the wind stress is based on a different blended wind speed product, mentioned in Section 2.1.  The admixture fractions are constrained so that they add up to 1 and are all between 0 and 1.  They are drawn randomly from a Dirichlet distribution.  Given limited observations, the wind products represent differences in interpolation methods and models \citep{kal96,gib97,com11,atl96}.  
\section{Results and Discussion}
\subsection{Is there a way to get around the uncertainty contributed by the wind?}
We approach this question by breaking it down into three parts: 1) First we show that the ocean responds differently to parameter and wind perturbations by comparing climatological averages (2005-2007) of the ocean state with a reference experiment for two model evaluations as representative of the ensembles of parameter and blended wind experiments, for pure NASA winds (Exp 3) vs the default and for $Ri_c$=0.45 (Exp 5) vs $Ri_c=0.15$ (Exp 4).  Second, we argue that parameter and wind EOFs that are representative of those anomaly patterns and are defined in space on the sparse TOGA/TAO mooring array locations differ from one another.  This provides evidence that projecting model data differences onto the parameter EOFs (as we do in our test statistic) filters for the part of the signal that is sensitive to parameter perturbations.  Finally, we demonstrate that the variability in our test statistic value, to which we have applied the parameter EOF filter, is higher for the set of parameter experiments than it is for wind experiments.  
 
\subsubsection{Ocean State Sensitivities in Parameter and Wind Experiments}
The critical bulk Richardson number and NASA experiments chosen to show anomaly fields of climatological ocean state fields are representative of other similar wind or parameter perturbation experiments, at the same time being end members that show maximum anomalies.  In general, the most striking difference between NASA or ECMWF and NCEP reanalyses is that NASA and ECMWF have stronger trade wind across the domain (a relationship that is robust over time).  Increasing $Ri_c$ directly increases the intensity of turbulent mixing almost everywhere and at most times.

The overall effect of increasing $Ri_c$ will be to increase the boundary layer depth $h$ and the eddy diffusivity/viscosity above $h$ by proportionality, resulting in deeper mixed layers.  The main effect of increasing the winds is to increase momentum imparted to the ocean as well as the intensity of turbulent mixing.  To understand how the Equatorial undercurrent is affected when the trade winds are increased, it will be necessary to consider the corresponding heat flux anomalies relative to the default in our ocean only model as well.   

\begin{figure}[!h]
\begin{center}
\includegraphics[width=\textwidth]{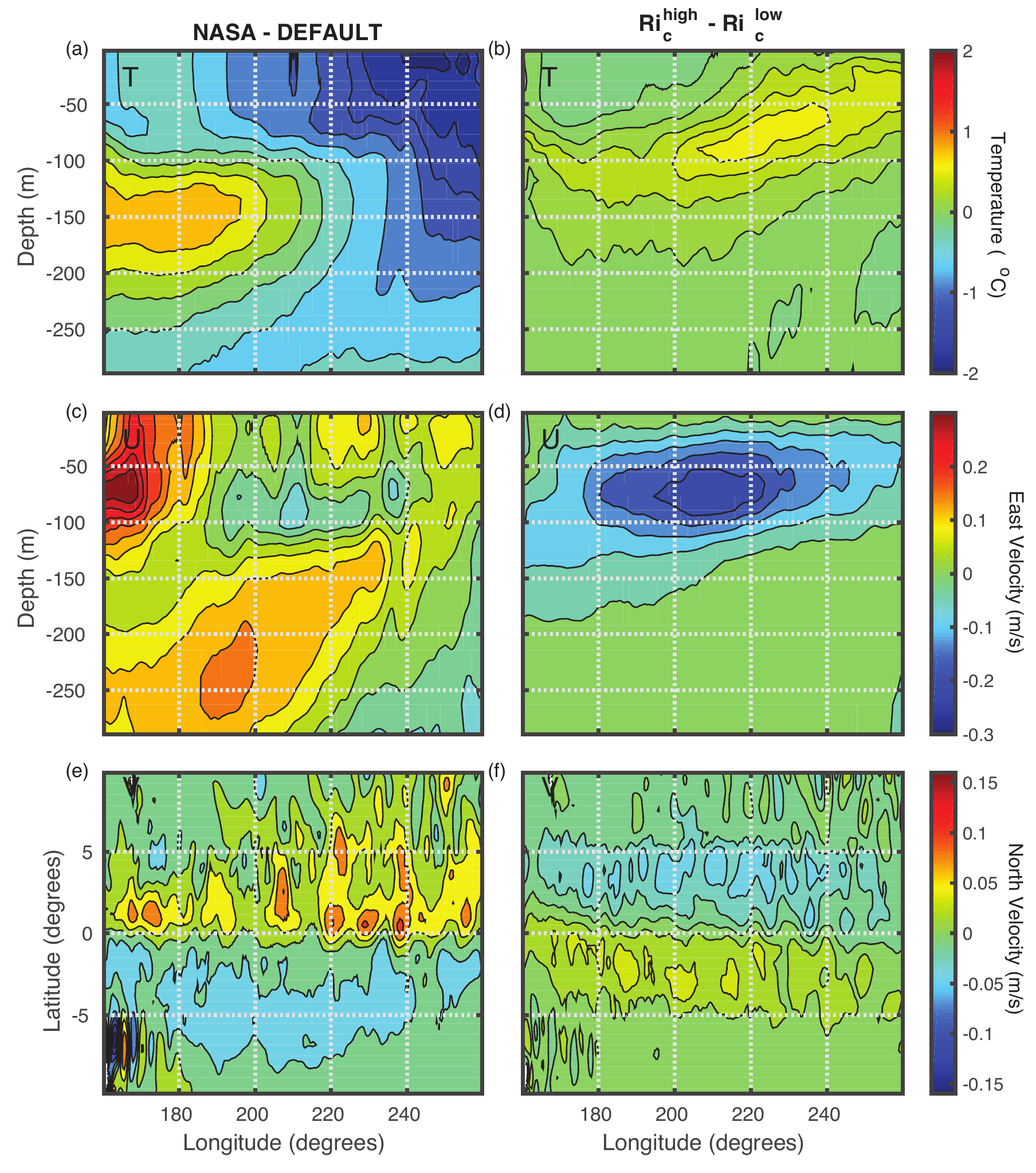}
\caption{
         Longitude-depth snapshots of Equatorial annually (2005-2007) averaged anomaly fields for temperature (a-b),
         east velocity (c-d) and north velocity (e-f)
         Left column.  NASA (Exp 3) - DEFAULT.  Right Column $Ri_c$=0.45 - $Ri_c$=0.15 (Exp 5 - Exp 4)}
\label{plot_Usens}
\end{center}
\end{figure}

When the critical bulk Richardson number is increased, the associated deeper mixed layers cause a reduction in speed of the the wind-forced Ekman currents to the north and south of the Equator  (Fig. \ref{plot_Usens}f; because turbulent mixing is more vigorous, and the imparted momentum is distributed over a thicker layer of water).  The fact that the anomaly field does not integrate to zero with depth is an artifact of the restorative surface heat and wind stress fluxes.  The weaker Ekman currents drive a weakened divergence at the Equator, which in turn reduces the vertical velocity and the upwelling of the thermocline there.
%%%%%%%%%%%%%%%%%%%%%%%%%%%%%%%%%%%%%%%%%%%%%%%%%%%%%%%%%%%%%%%%%%%%%%%%%%%%%%%%%%%%%%%%%%%%%%%%%%%%%%%%%%%%%%%%%%%%%%%%%%
%%%%%%%%%%%%%%%%%%%%%%%%%%%%%%%%%%%%%%%%%%%%%%%%%%%%%%%%%%%%%%%%%%%%%%%%%%%%%%%%%%%%%%%%%%%%%%%%%%%%%%%%%%%%%%%%%%%%%%%%%%

The combination of enhanced mixing and reduced Ekman upwelling at the Equator results in increased temperatures at the surface and in the thermocline in the eastern part of the basin.  In the west, the mixed layer cools and the thermocline warms, but by smaller amplitudes than in the east, because upwelling is weaker in the west and mixed layers are deeper (Fig. \ref{plot_Usens}b; so that any turbulent mixing induced changes in temperature are averaged over a thicker layer of water and hence are smaller). 

Because the equatorial current (EC) and equatorial undercurrent (EUC) flow in opposite directions, intensifying the mixing of eastward momentum across the sheared boundary between them will act to decrease both simultaneously (Fig. \ref{plot_Usens}d).  The maximum effect is in the central part of the basin, where surface wind stress forcing and convection are largest. 

When NASA and ECMWF reanalysis winds are blended into NCEP, the increased wind stress forces larger Ekman currents and a larger Ekman divergence at the Equator across the basin (Fig. \ref{plot_Usens}e).  On average, the increase in upwelling is slightly larger in the eastern than in the western part of the domain.

The mixed layer cools across the basin from a combination of enhanced upwelling and turbulence (Fig. \ref{plot_Usens}a)  The effect of increased upwelling on the temperature field is strongest in the east, where it is more important than that of vertical mixing.  In the west, the thermocline warms and the mixed layer cools, indicating that the effect of mixing is more important than that of upwelling there. 

The anomaly pattern is slightly more complicated for east velocity than it is for north velocity and temperature, but a key feature is that the maximum anomaly is in the western part of the domain (Fig. \ref{plot_Usens}c).  Overall, this results because of the character of the anomalies between NASA and NCEP wind stress and the corresponding heat flux forcing varies as a function of longitude along the Equator.  As shown at four locations along the Equator in Table \ref{qnet} the 
%%%%%%%%%%%%%%%%%%%%%%%%%%%%%%%%%%%%%%%%%%%%%%%%%%%%%%%%%%%%%%%%%%%%%%%%%%%%%%%%%%%%%%%%%%%%%%%%%%%%%%%%%%%%%%%%%%%%%%%%%%
\begin{table}[tp]
\begin{center}
\begin{tabular}[t]{|c|c|c|c|c|}
\hline
$\frac{\bigtriangleup \tau}{\tau^d}$ & 0.50 & 0.37 & 0.34 & 0.17 \\ \hline
$\frac{\bigtriangleup Q_{net}}{Q^d_{net}}$ & 0.07 & -0.07 & -0.31 & -0.48  \\ \hline
\end{tabular}
\end{center}
\caption{Fractional difference in NASA climatological averages of wind stress and net heat flux release to atmosphere relative to default NCEP values, at night and for the 2005 - 2007 period.}   
\label{qnet}
\end{table}
%%%%%%%%%%%%%%%%%%%%%%%%%%%%%%%%%%%%%%%%%%%%%%%%%%%%%%%%%%%%%%%%%%%%%%%%%%%%%%%%%%%%%%%%%%%%%%%%%%%%%%%%%%%%%%%%%%%%%%%%%%
largest mean fractional change in the Equatorial wind stress (2005-2007) relative to the default value is in the west, and decreases steadily to the east, whereas the largest mean fractional change in heat flux loss to the atmosphere is largest in the east, and decreases steadily to the west.  This can be reasoned out by considering the temperature anomaly pattern and the dependence of the bulk formulae in our ocean only model on temperature.  The large decrease in temperature in the east causes a large decrease in latent, sensible, and longwave heat fluxes to the atmosphere, despite the larger winds there (which have the opposite effect on the latent and sensible heat fluxes to the atmosphere).  In the east, the large reduction in heat flux loss to the atmosphere reduces the amount of convective mixing in the KPP, which partially compensates the effect of the modest increase in wind stress.  Overall, the EUC is increased.    In the western part of the domain, the heat fluxes are not changed much, but the wind stress changes are large, so the effect of the winds on the east velocity are not compensated by a reduction in convective mixing, and the changes in the EC/EUC are largest there.  

\subsubsection{Parameter and Wind EOFs are different}
In Fig. \ref{plot_Usens} we show difference fields for specific experiments, but we have 40 model evaluations to work with, and we have calculated a set of EOFs that represent the spatial anomaly patterns for the ensemble of 20 parameter experiments and that of 20 wind experiments, following the same procedure.  The anomaly patterns shown in Fig. \ref{plot_Usens} will be similar to the EOFs if they are robust for all experiments within the ensemble and if the EOFs, which exist on the array of TOGA/TAO sensor locations that is sparse in space, adequately represent the distinguishing features of those anomaly patterns.  The geometrical projection of a parameter EOF onto a wind EOF is a measure of how similar the patterns contained in the EOFS are, with values of 0 indicating they are orthogonal to one another and a value of 1 that they are parallel.  We calculated the projection all possible pairs of parameter and wind EOFs.  The mean and maximum projection values are presented in Table \ref{eof}.
%%%%%%%%%%%%%%%%%%%%%%%%%%%%%%%%%%%%%%%%%%%%%%%%%%%%%%%%%%%%%%%%%%%%%%%%%%%%%%%%%%%%%%%%%%%%%%%%%%%%%%%%%%%%%%%%%%%%%%%%%%
\begin{table}[tp]
\begin{center}
\begin{tabular}[t]{|c|c|c|}
\hline
Variable & Mean & Max \\ \hline
East velocity & 0.24 & 0.64 \\ \hline
North velocity & 0.21 & 0.88 \\ \hline
Temperature & 0.16 & 0.57 \\ \hline
Salinity & 0.16 & 0.61 \\ \hline
\end{tabular}
\end{center}
\caption{Mean and maximum projections of all possible pairs of wind and parameter EOFs, for each ocean state variable.} 
\label{eof}
\end{table}
%%%%%%%%%%%%%%%%%%%%%%%%%%%%%%%%%%%%%%%%%%%%%%%%%%%%%%%%%%%%%%%%%%%%%%%%%%%%%%%%%%%%%%%%%%%%%%%%%%%%%%%%%%%%%%%%%%%%%%%%%%

The mean projection of wind onto parameter EOFs is between 0.16 and 0.24 for all ocean state variables, suggesting that on average, the parameter and wind EOFs are mostly orthogonal to one another.  The largest maximum projection is for north velocity, with a value of 0.88.  It occurs for the first wind and first parameter EOF.  This is consistent with what is shown in Fig. \ref{plot_Usens}e-f, which are quite similar to one another.   For temperature, east velocity and salinity, the maximum projection ranges between 0.57 and 0.64, which is smaller, suggesting that the parameter and wind EOFs are somewhat different from one another in support of the hypothesis that their effects can be partially separated.  However, the question of whether our test statistic effectively filters for the part of the signal that is most sensitive to parameter perturbations has not yet been answered.   	

\subsubsection{Test Statistic Filters for Parameter Response}
\begin{figure}[!h]
\begin{center}
\includegraphics[width=\textwidth,height=6.0in]{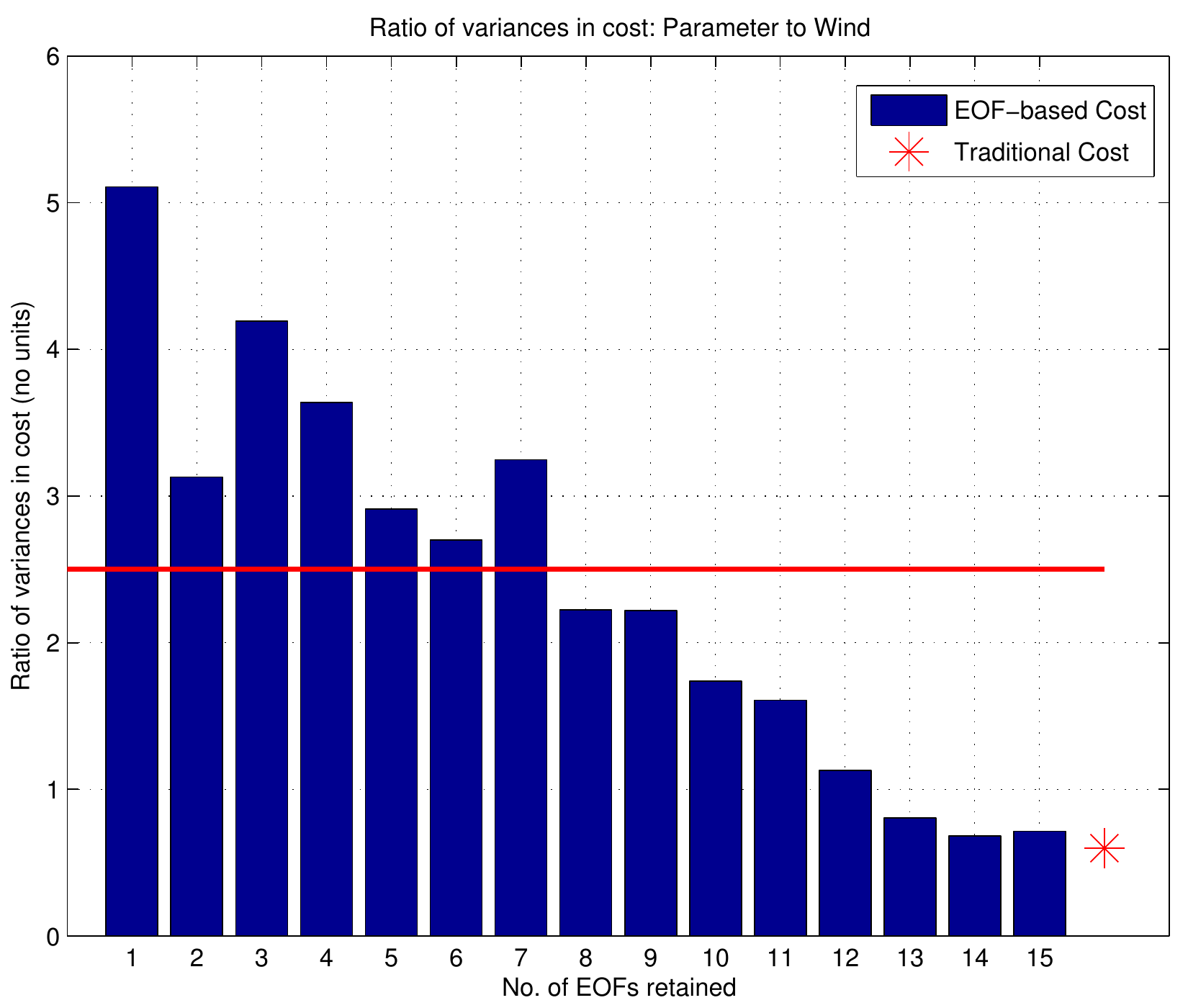}
\caption{The ratio of the variances in the ID test statistic values of the 
20 single KPP parameter perturbation experiments (including the default) 
 to the 20 blended wind experiments.
Using this figure we determine that a maximum of 7 EOFs should be retained in our test statistic.
For reference, we show the line with ratio of the variances equal to 2.5 in red.
}
\label{plot_ratio}
\end{center}
\end{figure}

As a test of the ability of our test statistic to filter out part of the wind forcing uncertainty, we turn again to the 20 wind experiments and 20 parameter experiments.  We calculated the test statistic value for each experiment, for different versions of the test statistic where the number of EOFs retained in the sum is varied from 1 to 15.  The ratio of variance in the test statistic for the 20 parameter perturbation experiments to that in the 20 wind experiments can be thought of as a relative gauge of the sensitivity of the test statistic to parameter perturbations.  Because it is difficult to determine whether the perturbations that were applied for parameters and winds are comparable, the actual magnitude of the ratio is irrelevant.  The important result is that the ratio decreases as a function of increasing retained EOFs in the sum, indicating that a larger fraction of the wind signal bleeds into the test statistic when a larger number of EOFs is retained, as shown in Fig. \ref{plot_ratio}.  This indicates 
%%%%%%%%%%%%%%%%%%%%%%%%%%%%%%%%%%%%%%%%%%%%%%%%%%%%%%%%%%%%%%%%%%%%%%%%%%%%%%%%%%%%%%%%%%%%%%%%%%%%%%%%%%%%%%%%%%%%%%%%%%
%%%%%%%%%%%%%%%%%%%%%%%%%%%%%%%%%%%%%%%%%%%%%%%%%%%%%%%%%%%%%%%%%%%%%%%%%%%%%%%%%%%%%%%%%%%%%%%%%%%%%%%%%%%%%%%%%%%%%%%%%%
that the filter for KPP parameter perturbations is more effective when small numbers of EOFs are retained in the test statistic.   This makes sense because the lower numbered EOFs contain the large spatial scales that represent the dominant patterns of sensitivity to parameter perturbations that contain most of the variance.   Small scale features will not be well represented in our coarse-scale model.  

In response to our first question, one may significantly reduce the effects of wind uncertainties in a test statistic by using an ID method to filter observations according to structures that matter most to parameters.

\subsection{How important is it that we synthesize all available data?}
\begin{figure}[!h]
\begin{center}
\includegraphics[width=\textwidth,height=6.0in]{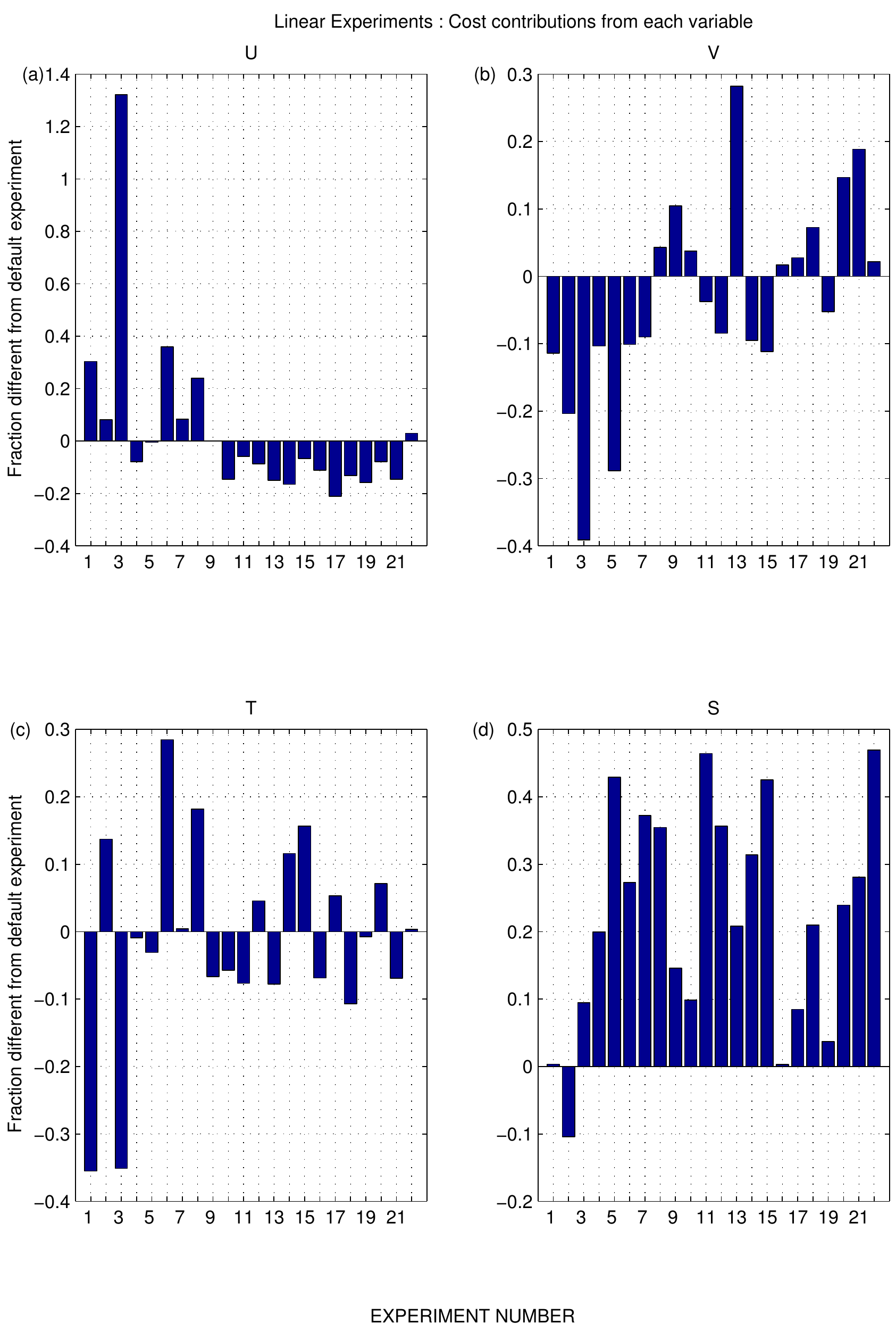}
\caption{Contributions to ID test statistic from each ocean state variable, presented as fraction different from default value: 
(a) for east component of velocity (U), (b) for north component of velocity (V), (c) for temperature (T), and (d) for salinity (S).
}
\label{plot_cost_var}
\end{center}
\end{figure}

The importance of synthesizing all available data will depend on the question one is asking.   In this case, we are
calibrating a turbulence model.  All ocean state variables are affected by turbulent mixing.  We have included them all in our test statistic, and if it is broken down by contributing components from each ocean state variable, it becomes clear that in a model calibration exercise, the optimal values and their uncertainty will depend on what data is included (Fig. \ref{plot_cost_var}).  
%%%%%%%%%%%%%%%%%%%%%%%%%%%%%%%%%%%%%%%%%%%%%%%%%%%%%%%%%%%%%%%%%%%%%%%%%%%%%%%%%%%%%%%%%%%%%%%%%%%%%%%%%%%%%%%%%%%%%%%%%%
%%%%%%%%%%%%%%%%%%%%%%%%%%%%%%%%%%%%%%%%%%%%%%%%%%%%%%%%%%%%%%%%%%%%%%%%%%%%%%%%%%%%%%%%%%%%%%%%%%%%%%%%%%%%%%%%%%%%%%%%%%
Different ocean state variables will have good matches to the data in different parts of parameter space.  By including all 
of them, our test statistic will be low in the regions of parameter space where ocean state variables tend to agree on a 
good match between model and data.  
\subsection{What is the potential to make use of moored buoy observations for calibrating uncertain parameters within the KPP?}
\begin{figure}[!h]
\begin{center}
\includegraphics[width=\textwidth,height=6.0in]{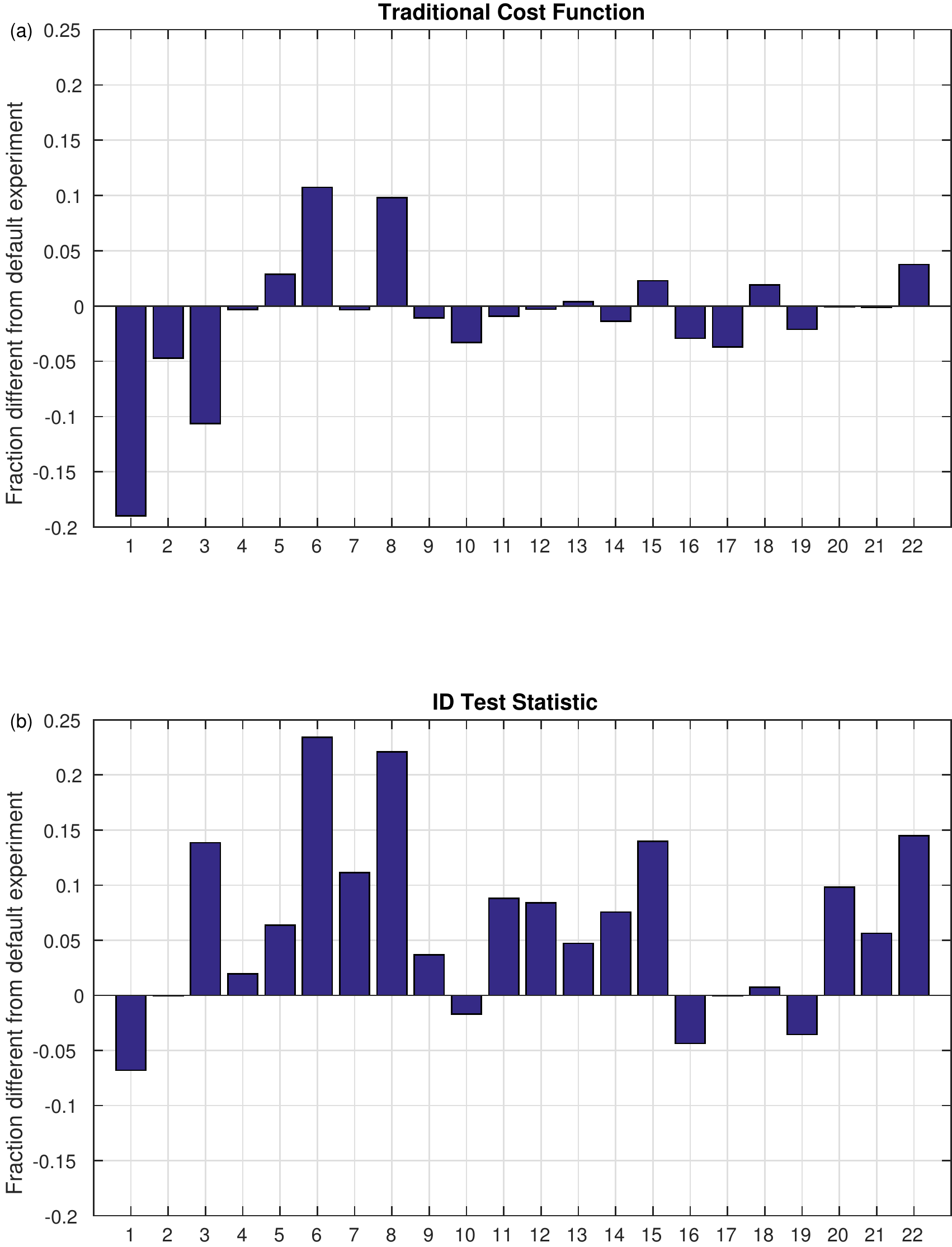}
\caption{ Fractional difference in test statistics from default value as a function of experiment number (from Table\ref{tab:linexpt}) values. (a) For simple cost function. (b)  For the ID test statistic. 
}
\label{plot_cost_tot}
\end{center}
\end{figure}

To demonstrate that our method of test statistic design is worth the effort, we compare it with a simple metric for 23 of the 40 experiments.  Our ID test statistic has more variability for the 19 single parameter perturbation experiments -- by a factor of 2 or more -- than the simple metric (Fig. \ref{plot_cost_tot}).  Both test statistics have about equal variability for the 23 wind experiments (20 blended not shown).  This suggests that the ID test statistic does a better job of distinguishing effects of parameters from the uncertainty in the wind forcing, and that it is appropriate for use in a Bayesian calibration of the KPP model against observations. 
%%%%%%%%%%%%%%%%%%%%%%%%%%%%%%%%%%%%%%%%%%%%%%%%%%%%%%%%%%%%%%%%%%%%%%%%%%%%%%%%%%%%%%%%%%%%%%%%%%%%%%%%%%%%%%%%%%%%%%%%%%
%%%%%%%%%%%%%%%%%%%%%%%%%%%%%%%%%%%%%%%%%%%%%%%%%%%%%%%%%%%%%%%%%%%%%%%%%%%%%%%%%%%%%%%%%%%%%%%%%%%%%%%%%%%%%%%%%%%%%%%%%%\section{Summary}
We developed a methodology of comparing model to data that takes into consideration processes, data, and sources of 
uncertainty that can affect the success of our ability to make use of the data to constrain uncertain parameters.  The method is developed in the context of a common turbulent mixing model, the K-Profile Parameterization as implemented in 
a coarse scale (1/3 degree) regional version of the MIT general circulation model configured for the tropical Pacific.  The inquiry-dependent test statistic of model data comparison filters model data discrepancies due to highly uncertain wind forcing.  The ID test statistic also accounts for spatial, temporal, and cross-correlations among different ocean state variables. 

The test statistic was then used to answer three questions that address some common challenges that arise in model data 
comparisons when the goal is to calibrate a turbulent mixing model.  First, we demonstrated that we can partially separate wind and parameter effects in our test statistic.  This is achieved by using the model as a proxy for the variability and relationships in the data and building a filter for the part of the data that is relevant to turbulent mixing.  Second, we showed that whether all the data should be included will depend in large part on the question one is asking; in a model calibration, the optimal parameter values and their uncertainty will depend on what data are included.  Turbulent mixing affects all ocean state variables, and including only a subset of the ocean state variables would run the risk of fitting one variable at the expense of another.  Therefore, it is important to include all data in a turbulent mixing calibration.  Finally, we conclude that moored buoy observations are useful for calibrating uncertain parameters within the KPP using our ID test statistic methodology.   In particular, we showed that our test statistic is more sensitive to parameter perturbations than a simple test statistic based on a sum of squared and variance normalized model data discrepancies.  Overall, this is a small step towards the broader goal of performing parameter calibrations to build more predictive models. 

\section{Acknowledgments}
We would like to acknowledge KAUST (KAUST grant CRG-1-2012-HOT-007 and AEA grant to UT Austin) and NASA who contributed through a subcontract with Boston University and the Massachusetts Institution of Technology.  This research made use of the resources of the Supercomputing Laboratory at King Abdullah University of Science and Technology (KAUST) in Thuwal, Saudi Arabia.  We would especially like to acknowledge Andrew Winfer, Ian Shore, and the team of system administrators who have been very helpful for this project in many ways, not least of which excellent maintenance of the KAUST machine Shaheen, the parallel computer platform that was used to run the model.

\bibliographystyle{model2-names}
\bibliography{library_2}

%% not want to use model2-names.bst.



%% \bibitem must have one of the following forms:
%%   \bibitem[Jones et al.(1990)]{key}...
%%   \bibitem[Jones et al.(1990)Jones, Baker, and Williams]{key}...
%%   \bibitem[Jones et al., 1990]{key}...
%%   \bibitem[\protect\citeauthoryear{Jones, Baker, and Williams}{Jones
%%       et al.}{1990}]{key}...
%%   \bibitem[\protect\citeauthoryear{Jones et al.}{1990}]{key}...
%%   \bibitem[\protect\astroncite{Jones et al.}{1990}]{key}...
%%   \bibitem[\protect\citename{Jones et al., }1990]{key}...
%%   \harvarditem[Jones et al.]{Jones, Baker, and Williams}{1990}{key}...
%%


% \end{thebibliography}
%\end{linenumbers}
\end{document}